\documentstyle[twoside,12pt]{article}
\normalsize
\newcommand{\bdi}{\begin{displaymath}}
\newcommand{\edi}{\end{displaymath}}
\newcommand{\bfi}{\begin{figure}}
\newcommand{\efi}{\end{figure}}

\newcommand{\beq}{\begin{equation}}
\newcommand{\eeq}{\end{equation}}

\newcommand{\gam}{\gamma_{\mu}}

\newcommand{\dsla}{\partial\hspace{-6pt} /  }  
\newcommand{\Asla}{A\hspace{-6.5pt}  /  } 

\newcommand{\psla}{p\hspace{-5.375pt} /   }

\newcommand{\wt}{\widetilde}

\begin{document}
\begin{titlepage}
\begin{flushright}
\today
\end{flushright}

\vspace{1cm}
\begin{center}
{\Large \bf Normalization of the chiral condensate in the massive Schwinger 
model}\\[1cm]
C. Adam* \\
School of Mathematics, Trinity College, Dublin 2 \\

\vfill
{\bf Abstract} \\
\end{center}

Within mass perturbation theory, already the first order contribution to the
chiral condensate of the massive Schwinger model is UV divergent. 
We discuss the problem of choosing a proper
normalization and, by making use of some bosonization results, we are able to
choose a normalization so that the resulting chiral condensate may be compared,
e.g., with lattice data.

\vfill

$^*)${\footnotesize  
email address: adam@maths.tcd.ie, adam@pap.univie.ac.at}
\end{titlepage}

\section{Introduction}

The massive Schwinger model, or massive QED$_2$ with one fermion flavour,
\beq
L = \bar\Psi (i\dsla -e\Asla -m)\Psi -\frac{1}{4}F_{\mu\nu}F^{\mu\nu},
\eeq
has been studied for some time because it resembles QCD in many respects.
The investigation started more than 20 years ago with some classical
papers \cite{CJS} -- \cite{FS1} and has continued ever since (for a review see
e.g. \cite{AAR,MSMPT}). Quite recently, the model has been studied in some
detail within mass perturbation theory \cite{MSMPT} -- \cite{SCAT}
as well as with light-front methods \cite{Heinzl1} -- \cite{Burk1},
on the lattice \cite{CaKe1} -- \cite{FHTS}, and by a generalized Hartree-Fock
approach on the circle \cite{Hoso1,HoRo1}. The multi-flavour case, too, has
received some attention recently \cite{HoRo1} -- \cite{Gatt2}.

Some of the features that make the model so attractive are the presence of 
instanton-like gauge field configurations, and, consequently, a nontrivial
vacuum structure ($\theta$ vacuum) \cite{LS1,CJS,Co1}; 
further the chiral anomaly  and the 
formation of a nonzero chiral condensate $\langle \bar\Psi \Psi\rangle$
\cite{LS1}.

In the massless case ($m=0$), the chiral condensate may be computed exactly
\cite{LS1}, \cite{HeHo1} -- \cite{Adam}.
For $m\ne 0$, a mass perturbation expansion can be performed and corrections
to $\langle \bar\Psi \Psi\rangle_{m=0}$ may be computed \cite{MSSM,MSMPT}. 
However, already at 
order $m^1$ a UV singularity occurs that has to be regularized, and a proper 
normalization for  $\langle \bar\Psi \Psi\rangle_{m}$ has to be chosen. It
is the purpose of this article to discuss this point and to arrive at an
expression for $\langle \bar\Psi \Psi\rangle_{m}$ that may be compared, e.g.,
to lattice computations. In the sequel, all computations are done for
two-dimensional, Euclidean space-time.

\section{Chiral condensate up to order $m$}

The Euclidean, bosonized version of the theory (1) reads \cite{CJS,Co1,FS1,
Gatt1}
\beq
L_{\rm b}= -N_\mu \Bigl[\frac{1}{2} \phi (\Box -\mu^2)\phi +
\frac{e^\gamma}{2\pi}\mu m
\cos (\sqrt{4\pi}\phi +\theta)\Bigr]
\eeq
where $\mu =e/\pi^{1/2}$ is the Schwinger mass, $N_\mu$ denotes normal
ordering w.r.t. $\mu$, $\theta$ is the vacuum angle, and $\gamma =0.5772$ 
is the Euler constant.

The vacuum condensate $\langle \bar\Psi \Psi\rangle_{m}$ is 
\beq
\langle \bar\Psi \Psi\rangle_{m} =-\frac{e^\gamma}{2\pi}\mu
\langle \cos (\sqrt{4\pi}\phi + \theta)\rangle_m
\eeq
within the bosonized version of the theory and may be evaluated by a
perturbation expansion in $m$. The lowest order expression is the wellknown
condensate of the massless model \cite{Jay} -- \cite{Adam}
\beq
\langle \bar\Psi \Psi\rangle_{m=0} =-\frac{e^\gamma}{2\pi}\mu\cos\theta .
\eeq
For a higher order computation it is useful to rewrite the interaction
part of the bosonic Lagrangian like
\beq
L_{\rm b,I}= -\frac{e^\gamma}{2\pi}\mu m \frac{1}{2} N_\mu \Bigl[ e^{i\theta}
e^{i\sqrt{4\pi}\phi} + e^{-i\theta}e^{-i\sqrt{4\pi}\phi}\Bigr] ,
\eeq
because the exponentials $\exp (\pm i\sqrt{4\pi}\phi)$ have especially simple
$n$-point functions within the massless model, e.g. (see \cite{MSMPT})
\beq
\langle e^{\sigma_1 i\sqrt{4\pi}\phi(x_1)}e^{\sigma_2 i\sqrt{4\pi}\phi(x_2)}
\rangle_{m=0} =e^{\sigma_1 \sigma_2 4\pi D_\mu (x_1 -x_2)}
\eeq
where $\sigma_1 ,\sigma_2 =\pm 1$, $D_\mu (x)$ is the massive scalar 
propagator
\beq
D_\mu (x)=-\frac{1}{2\pi}K_0 (\mu |x|), \qquad \wt D_\mu (p)=
\frac{-1}{p^2 +\mu^2}
\eeq
($K_0$ \ldots \, McDonald function). Further, powers of $e^{i\theta}$ indicate
the contributing instanton sectors ($e^{in\theta}\, \ldots \;$ instanton number
$k=n$). 

The exponentials $\exp (\pm 4\pi D_\mu (x))$ have the limit $\lim_{|x| 
\to\infty} \exp (\pm 4\pi D_\mu (x)) = 1$, therefore a disconnected piece
has to be subtracted, and the correct propagators for the mass perturbation 
expansion are the functions
\beq
E_\pm (x,\mu)=e^{\pm 4\pi D_\mu (x)} -1 .
\eeq
For the chiral condensate in order $m^1$ one finds easily 
(see \cite{MSSM,MSMPT}; in \cite{MSSM,MSMPT} the sign of $\langle\bar\Psi
\Psi\rangle$ is reversed due to an opposite sign convention for the
fermion mass $m$ in the Lagrangian)
\bdi
\langle \bar\Psi \Psi\rangle^{(1)}_m =-\frac{1}{2}m\mu^2 \frac{e^{2\gamma}}{
4\pi^2}\int d^2 x\bigl( E_+ (x,\mu)\cos 2\theta + E_- (x,\mu)\bigr)
\edi
\beq
=-\frac{1}{2}m\frac{e^{2\gamma}}{4\pi^2}\int d^2 x\bigl( E_+ (x,1)\cos 2\theta
+E_- (x,1)\bigr) .
\eeq
Taking into account the short-distance behaviour of $K_0 (z)$,
\beq
K_0 (z) \sim -\gamma -\ln\frac{z}{2} \qquad {\rm for} \qquad z\to 0
\eeq
one easily finds that the integral w.r.t. $E_+ (x,1)$ is UV finite,
\beq
E_+ := \int d^2 xE_+ (x,1)=-8.9139
\eeq
whereas the integral w.r.t. $E_- (x,1)$ is logarithmically UV divergent. 
Actually, these findings can be understood immediately from ordinary 
perturbation theory (in $e$) for the massless Schwinger model. The $E_+
(x,1)$ contribution in (9) is purely non-perturbative (in $e$) (it receives
contributions from the instanton sectors $k=\pm 2$). On the other hand, the
$E_- (x,1)$ contribution (from the $k=0$ sector)
\beq
\langle\bar\Psi\Psi\rangle^{(1),k=0}_m =-\frac{m}{2}\int d^2 x\langle
\bar\Psi (x)\Psi (x)\bar\Psi (0)\Psi (0)\rangle^{k=0}_{{\rm c},m=0}
\eeq
(c $\ldots \;$ connected part) is purely perturbative. For small 
distances $|x|$ the integrand in (12) behaves like
\beq
\langle\bar\Psi (x)\Psi (x)\bar\Psi (0)\Psi (0)\rangle^{k=0}_{{\rm c},m=0}
=\mu^2 \frac{e^{2\gamma}}{4\pi^2}e^{2K_0 (\mu|x|)}-1 \sim\frac{1}{2\pi^2 x^2},
\eeq
which is just the lowest order contribution of the
perturbative expansion (in $e$) within the massless Schwinger model
\beq
\langle\bar\Psi (x)\Psi (x)\bar\Psi (0)\Psi (0)\rangle^{k=0}_{m=0}=
{\rm tr} G_0 (x)G_0 (-x) +o(e^2)=\frac{1}{2\pi x^2} +o(e^2)
\eeq
where $G_0 (x)=i\gam x^\mu /(2\pi x^2)$ is the free, massless fermion 
propagator. Higher order contributions may be ignored in the
$|x| \to 0$ limit (asymptotic freedom). [These higher order terms consist
of all possible insertions of massive photon lines into the two massless
fermion propagators $G_0 (x)$, and turn out to exponentiate, leading to formula
(6) for $\sigma_1 \sigma_2 =-1$ (the photon acquires the Schwinger mass $\mu$
via the Schwinger mechanism, i.e., the summation of all vacuum polarization
insertions, see \cite{Sc1,ABH,PERT}).]

In \cite{MSSM} we regulated the chiral condensate by just isolating this free
fermion singularity via a partial integration 
\bdi
\langle\bar\Psi \Psi\rangle^{(1),k=0}_m =-\frac{m}{2}\frac{e^{2\gamma}}{4\pi^2}
\int d^2 x(e^{2K_0 (|x|)}-1)
\edi
\bdi
=-\frac{m}{2}\frac{e^{2\gamma}}{4\pi^2}\lim_{\epsilon\to 0}2\pi
\int_\epsilon^\infty \frac{dr}{r}(e^{2K_0 (r)+2\ln r}-r^2)
\edi
\bdi
=-m\pi \frac{e^{2\gamma}}{4\pi^2}\Bigl( \, \lim_{\epsilon\to 0} \, \Bigl[ \ln r
(e^{2K_0 (r)+2\ln r}-r^2)\Bigr]_\epsilon^\infty
\edi
\bdi
+\int_0^\infty dr\ln r [2(K_1 (r)-\frac{1}{r})e^{2K_0 (r)+2\ln r}+r]\Bigr)
\edi
\beq
=\frac{m}{\pi} \lim_{\epsilon\to 0}\ln\epsilon -\frac{m}{2}
\frac{e^{2\gamma}}{4\pi^2}E_-
\eeq
\beq
E_- := 2\pi\int_0^\infty dr\ln r [2(K_1 (r)-\frac{1}{r})e^{2K_0 (r)+2\ln r}+r]
=9.7384
\eeq
($K_0' =-K_1$)
where we performed the limit where it is safe. Normalizing the chiral 
condensate by just omitting the $\ln\epsilon$ term is plausible from the 
viewpoint of mass perturbation theory, because the latter relies on the
(exact solution of the) massless Schwinger model, and omitting the 
$\ln\epsilon$ term just amounts to omitting the non-interacting contribution
of the massless Schwinger model.

However, this normalization is not appropriate for a comparison with, e.g., 
lattice data (as was pointed out in \cite{HoRo1}). Instead, one has to choose
the normalization of ordinary perturbation theory,
\beq
\langle\bar\Psi\Psi\rangle_m' =\langle\bar\Psi\Psi\rangle_m - 
\langle\bar\Psi\Psi\rangle^{e=0}_m .
\eeq
Here one may wonder whether this normalization may be chosen within the 
context of mass perturbation theory. We will find that this is possible due 
to the specific two-dimensional feature of bosonization, as we want to
discuss now.

To obtain $\langle\bar\Psi\Psi\rangle^{e=0}_m$
in the bosonic language, we would just like to redo our computation for the
$\mu\to 0$ limit of the bosonic Lagrangian $L_{\rm b}$, (2). However, as
both coupling constants and the normal ordering in $L_{\rm b}$ depend on
$\mu$, we should first do a renormal-ordering. Using the normal-ordering
relation (see \cite{Co2})
\beq
N_m e^{\pm i\beta\phi (x)}=(\frac{\mu}{m})^{\frac{\beta^2}{4\pi}}
N_\mu e^{\pm i\beta\phi (x)}
\eeq
we find (up to an irrelevant additive constant)
\beq
L_{\rm b}=-N_m \Bigl[ \frac{1}{2} \phi (\Box -\mu^2)\phi +
\frac{e^\gamma}{2\pi}m^2
\cos (\sqrt{4\pi}\phi +\theta)\Bigr]
\eeq
where now the limit $\mu\to 0$ can be performed. Further, for $\mu =0$
the vacuum angle $\theta$ may be compensated by a shift of the field $\phi$
and can, therefore, be set equal to zero (i.e., there are no instanton
sectors when there is no gauge interaction), and one gets
\beq
L_{\rm b}^{e=0} =-N_m \Bigl[ \frac{1}{2} \phi\Box\phi +
\frac{e^\gamma}{2\pi}m^2
\cos\sqrt{4\pi}\phi \Bigr] .
\eeq
Within the bosonic approach we now would be left with the task of performing
a perturbation expansion for a massless scalar field, which is IR
divergent. But at this point bosonization results may be used. The Lagrangian
$L_{\rm b}^{e=0}$ is a version of the sine-Gordon model, which is known to
be the QFT analog of the massive Thirring model \cite{Co2}. More precisely,
after a volume cutoff is introduced, the perturbative expansion of the 
sine-Gordon model with the cosine term as interaction Lagrangian is
equivalent to a perturbative expansion of the massive Thirring model with
the fermion mass term as interaction term. Specifically, when the 
coefficient $\beta$ in $\cos\beta\phi$ is $\beta =\sqrt{4\pi}$, the
bosonic theory (20) is equivalent to the ``Thirring model'' with zero
coupling, i.e., the model with one free, massive fermion. The equivalence
of the two perturbation expansions mentioned above implies that the 
fermionic Lagrangian has to be normal-ordered w.r.t. zero fermion mass. 

The essential point is, of course, that the VEV $\langle\bar\Psi\Psi
\rangle_m^{e=0}$ can be computed exactly in the fermionic formulation, 
without the need to actually perform a (IR-divergent) mass perturbation
expansion. The correct normal-ordering prescription just means that the
regularized expression for $\langle\bar\Psi\Psi\rangle_m^{e=0}$ has to
vanish in the $m\to 0$ limit. Explicitly we find from ordinary Feynman rules
\beq
\langle\bar\Psi\Psi\rangle_m^{e=0}=\int\frac{d^2 p}{4\pi^2}{\rm tr}\frac{-i}{
\psla -im}=\int\frac{d^2 p}{4\pi^2}\frac{2m}{p^2 -1}=
-\frac{m}{4\pi^3}\int d^2 pd^2 xe^{ipx}K_0 (|x|) .
\eeq
Once the integral is regularized, this expression indeed vanishes for $m\to 0$.

Now we have to subtract this expression from the $k=0$ contribution to the
chiral condensate in order $m$, (9). For a unified regularization prescription 
of both terms we slightly shift the argument of $K_0 (|x|)$, $K_0 (|x|)\to
K_0 (|x| +\epsilon)$, in both expressions and obtain 
\bdi
\langle \bar\Psi\Psi\rangle^{'(1),k=0}_m =
\langle \bar\Psi\Psi\rangle^{(1),k=0}_m - 
\langle \bar\Psi\Psi\rangle^{e=0}_m 
\edi
\bdi
= -\frac{m}{2}\frac{e^{2\gamma}}{4\pi^2}2\pi \int_0^\infty drr (e^{2K_0 (r+
\epsilon)}-1) +\frac{m}{\pi}\int d^2 x\delta (x)K_0 (x+\epsilon)
\edi
\bdi
= -m\pi\frac{e^{2\gamma}}{4\pi^2}\int_\epsilon^\infty dr(r-\epsilon)
(e^{2K_0 (r)}-1) +\frac{m}{\pi}K_0 (\epsilon)
\edi
\bdi
= -m\pi\frac{e^{2\gamma}}{4\pi^2}\int_\epsilon^\infty dr (\frac{1}{r}-
\frac{\epsilon}{r^2})(e^{2K_0 (r)+2\ln r} - r^2) + \frac{m}{\pi}K_0 (\epsilon)
\edi
\bdi
= \frac{m}{\pi}\ln\epsilon -\frac{m}{2}\frac{e^{2\gamma}}{4\pi^2}E_- +
\frac{m}{\pi}(-\ln\epsilon -\gamma +\ln 2)
\edi
\bdi
+ m\pi \frac{e^{2\gamma}}{4\pi^2}\epsilon \Bigl[ \frac{-1}{r}(e^{2K_0 (r)
+2\ln r}-r^2)\Bigr]_\epsilon^\infty +o(\epsilon)
\edi
\beq
=\frac{m}{\pi}(-\gamma +\ln 2) -\frac{m}{2}\frac{e^{2\gamma}}{4\pi^2}E_-
+m\pi\frac{e^{2\gamma}}{4\pi^2}e^{-2\gamma +2\ln 2} +o(\epsilon)
\eeq
where we performed a partial integration in both integrals and used the
result (15) for the first integral. The remainder of the second integral
that we did not display explicitly is of order $\epsilon$. Putting 
everything together, we obtain for the chiral condensate up to order
$m$
\beq
\langle\bar\Psi\Psi\rangle^{'}_m =-\frac{e^\gamma}{2\pi}\mu\cos\theta
-m\Bigl[ \frac{e^{2\gamma}}{8\pi^2}(E_+ \cos 2\theta +E_- )-\frac{1}{\pi}
(1+\ln 2 -\gamma )\Bigr]
\eeq
\beq
=-0.283 \, \mu\cos\theta +0.358 \, m\cos 2\theta -0.036 \, m
\eeq
where we inserted all the numbers in the last line. For $\theta =0$ we obtain
\beq
\langle\bar\Psi\Psi\rangle^{'}_{m,\theta =0}=-0.283 \, \mu +0.322 \, m
\eeq
which may be compared to the lattice results of \cite{FHTS} and to the 
generalized Hartree-Fock computation of \cite{HoRo1}, see Fig. 1. The results 
can be seen to agree modestly for sufficiently small fermion mass $m$.
\input psbox.tex 
\begin{figure}
$$\psbox{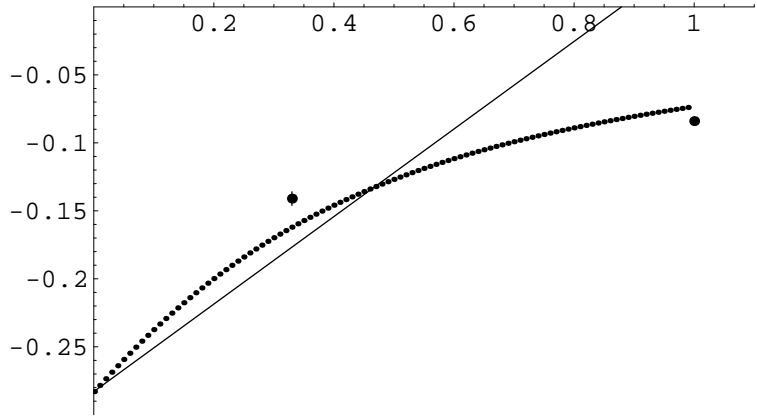}$$
\caption{The figure shows fermion mass $m/\mu$ ($x$-axis) vs. chiral
condensate $\langle\bar\Psi\Psi\rangle_m' /\mu$ ($y$-axis), both in units
of $\mu$. The straight line is the mass perturbation result (25); the
dotted curve is the generalized Hartree-Fock result of \cite{HoRo1};
the two points with error bars are the lattice results $(0.33,-0.141)$,
$(1.00,-0.084)$ of \cite{FHTS}}
\end{figure}

\section{Discussion}

With the help of bosonization arguments we succeeded in finding a 
normalization of the chiral condensate, $\langle\bar\Psi\Psi\rangle^{'}_m
=\langle\bar\Psi\Psi\rangle_m - \langle\bar\Psi\Psi\rangle^{e=0}_m$,
that implies a matching between mass perturbation theory and ordinary,
electric charge perturbation theory. This normalization is the appropriate one
for a comparison with lattice data.

Observe that we have two consistency checks for our bosonization prescription
of $\langle\bar\Psi\Psi\rangle^{e=0}_m$. Firstly, the precise cancellation 
of the singularity shows that we have properly matched the ``coupling
constants'' of the bosonic and fermionic prescriptions of 
$\langle\bar\Psi\Psi\rangle^{e=0}_m$. Secondly, as already mentioned,
the regularized expression for $\langle\bar\Psi\Psi\rangle^{e=0}_m$ has to
vanish in the $m\to 0$ limit because of the bosonization rules, which it
does indeed. Actually, $\langle\bar\Psi\Psi\rangle^{e=0}_m$ turns out to
be exactly of order $m$, (21), as it must be for dimensional reasons.

This fact may lead to some speculation about UV-finiteness. The point is
that the normalization $\langle\bar\Psi\Psi\rangle^{'}_m =
\langle\bar\Psi\Psi\rangle_m -\langle\bar\Psi\Psi\rangle^{e=0}_m$
is the normalization of ordinary (in $e$) perturbation theory. Further,
within ordinary perturbation theory, the massive Schwinger model is
super-renormalizable (i.e., finite after normal-ordering) \cite{CJS,Co1}. 
This leads to
the conjecture that all the higher order contributions to 
$\langle\bar\Psi\Psi\rangle_m$ within mass perturbation theory are UV
finite. It is, however, a rather difficult (and not yet solved) problem
to prove (or disprove) this UV finiteness directly within mass
perturbation theory.

\section*{Acknowledgement}

The author thanks the members of the Department of Mathematics at Trinity
College, where this work was performed, for their hospitality. Further
thanks are due to Y. Hosotani for the data of \cite{HoRo1} and
\cite{FHTS} in Fig. 1.

\end{document}